\documentclass{article}
\usepackage{booktabs,spconf,amsmath,graphicx,hyperref}
\usepackage{multirow}
\usepackage{indentfirst}
\usepackage{xcolor,url}

\newcommand{\lefour}{$\le$4}
\newcommand{\gt}[1]{$>$#1}
\title{Slots Are Not Speakers: Training-Free Speaker Linking for Frozen Streaming Diarizers}
\name{Aditya Srinivas Menon, Kumarmanas Nethil, Kavya Manohar}
\address{Adalat AI, India}
\begin{document}
%
\maketitle
%

\begin{abstract}
This paper presents a causal, training-free linker that allows a frozen streaming diarizer to track more speakers than it has output slots. Streaming end-to-end diarizers such as Streaming Sortformer emit a fixed number N of concurrent speaker slots. Above N speakers, they keep detecting speech but reuse slots across people, because N bounds how many speakers the model can express at once, not how many it will meet. The linker uses only the host's slot posteriors and the audio. It keeps the frames where one slot is active, splits each slot where the speaker changes, embeds each piece with an off-the-shelf speaker encoder, and matches it against a bank of speakers seen so far. It only relabels, so missed speech and false alarm are identical to the host's and every gain is a reduction in speaker confusion. On 1418 sessions from eight corpora, it reduces the 4-slot host's DER above its cap from 36.8\% to 22.6\%, and barely moves it below the cap. The same constants transfer to the host's 8-slot successor and to two streaming EEND hosts from a different model family. On courtroom hearings with up to 23 speakers it reduces cpWER from 82.7\% to 24.4\%.
\end{abstract}


\section{Introduction}
\label{sec:intro}

End-to-end neural diarizers emit a fixed number of output channels~\cite{fujita2019eend,horiguchi2020eda}, and their streaming variants fix that number when they are trained~\cite{streamsortformer,fseend,lseend}. Streaming Sortformer~\cite{streamsortformer, park2025sortformernovelapproachpermutationresolved} emits $N{=}4$ arrival-ordered speaker slots. Production ships the checkpoint that exists, and meetings do not respect $N$. On a 13-speaker earnings call~\cite{earnings21} (Fig.~\ref{fig:strip}) a 4-slot host does not fail loudly. It gets the voice activity regions right, but the 4 slots are reused across people, and the transcript silently merges speakers. Pooled over the 505 sessions in our suite with more than four speakers, its diarization error rate (DER) is 36.8\,\%, of which 25.3 points are speaker confusion.


The authors describe the model as tracking ``the four most dominant speakers in a conversation''~\cite{streamsortformer}. Fig.~\ref{fig:strip} shows slot~2 being occupied by 5 speakers for over 68 minutes. A slot is not a speaker but a lane, freed when its occupant falls silent and handed to whoever speaks next. The host's arrival-order speaker cache~\cite{streamsortformer} is not a persistent identity store (Sec.~\ref{sec:linker}), so a speaker who returns after their lane has been reassigned is admitted as new. $N$ is therefore a \emph{concurrency} cap, not a \emph{lifetime} cap. EEND-GLA~\cite{horiguchi2023gla} draws the same distinction for block-local attractors, whose total speaker count ``can be higher than the limitation'' within a block, and resolves it with a jointly trained linker.
 
The host already frees and reuses slots, but what it lacks is a global speaker identity memory that lives within the session. Identity can therefore be restored outside the model, if the linker is causal, cheap, and does not disturb the host where the host is already right. Pairing a capacity-limited local diarizer with a global linker is an established design~\cite{eendvc1,horiguchi2023gla,xue2021stb}, and diart~\cite{diart} does it online. This work adds a global linker to a frozen production host, with nothing trained and no frame of its activity altered. The linker relabels and does nothing else, so missed speech and false alarm are from the host by construction, and every change it makes is speaker confusion alone.
 
We report four results as our contribution. On 1418 sessions from eight corpora under one scorer, the linker cuts the DER of the 4-slot host~\cite{sortformerv21card} from 36.8 to 22.6 above its cap, and moves it by $-0.08$ below (paired bootstrap CI $[-0.41, +0.24]$). The same constants, unchanged, lower DER on the host's 8-slot successor~\cite{nemotron3card} above its higher cap and on two streaming EEND hosts from a different model family~\cite{fseend,lseend}, on the same sessions. Widening this host's head to eight slots by fine-tuning~\cite{ultrasortformer} does not close the gap: the frozen 4-slot host plus the linker (18.41) beats the widened host plus the same linker (21.42). On four courtroom hearings with up to 23 speakers, where the host predicts four speakers throughout, the linker recovers 20/23, 14/16 and 10/14 identities, and takes cpWER from 82.7 to 24.4.
 
\section{Training-free linker}
\label{sec:linker}

\textbf{Observation.} On the 4-slot host, pooled over 1418 sessions (213.5\,h, eight corpora), missed speech and false alarm barely change with speaker count. Below the cap (913 sessions) they are 8.65 and 2.61, above it (505 sessions) 8.93 and 2.59. Speaker confusion goes from 4.07 to 25.29, and DER with it, from 15.32 to 36.81. Above the cap the host still finds the speech. What it loses is who said it.

\textbf{Setting.} The frozen host emits at 12.5 frames per second and in causal chunks, a posterior tensor of shape $[T, N]$. At most $N$ concurrent slot activations. The linker consumes that tensor and the audio. It reads no cache, embedding or internal state, so it applies to any diarizer emitting per-channel activity posteriors, and the host is never modified.

\begin{figure}[t]
\centering
\includegraphics[width=\columnwidth,height=2.4in,keepaspectratio]{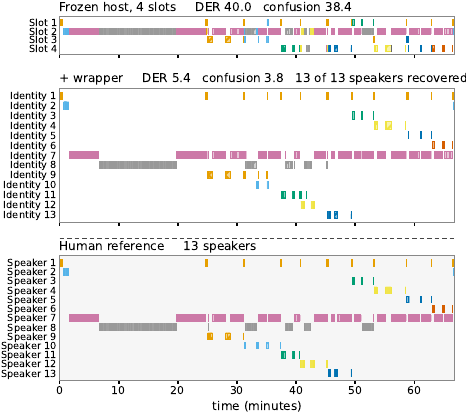}
\caption{Earnings-21 session 018, 13 speakers. \textbf{Top}: the frozen 4-slot host, regions coloured by dominant reference speaker; Slot~2 alone carries five people. \textbf{Middle}: the linker, 13 of 13 identities. \textbf{Bottom}: the reference. DER 40.0 to 5.4.
}
\label{fig:strip}
\end{figure}

 
\textbf{Pipeline.} (Fig.~\ref{fig:strip}). The linker has four steps and two rules, all causal. \emph{Select.} For every frame we compute a purity score, the log-odds from the slot posteriors that exactly one slot is active, and keep the purest 60\,\% of frames in each region. Overlapped and uncertain frames therefore never enter an embedding. \emph{Segment.} Each slot's timeline is cut into stretches at speaker change points: a new stretch starts when a region's embedding has cosine similarity below 0.40 to the running centroid of the current stretch. The stretch is the unit the linker decides on. \emph{Embed.} The kept frames of a stretch are embedded with TitaNet-large~\cite{titanet} (192-d, 25.3\,M parameters, run on CPU) and mean-pooled. A bank holds one centroid per speaker identity. \emph{Link.} The first stretch on a slot keeps that slot's identity. A later stretch takes the bank identity with the highest cosine similarity if that similarity is at least 0.60~\cite{diart}; moving a slot away from the identity it already carries needs a margin of 0.15. \emph{Cannot-link.} A stretch may not join an identity that was active on another slot for more than 2\,s of the same span: two slots speaking at once are two people~\cite{eendvc1,diart}. \emph{Fold-back.} The linker emits a new identity only after 15\,s of its speech. It adds no trainable parameters and runs on CPU beside the host at a real-time factor of 0.019.
 
\textbf{Invariant and bounds.} The linker only relabels, so missed speech and false alarm are the host's. A new identity is emitted only once it has 15\,s of speech or more than ten stretches, so a speaker who never reaches either stays under the slot label. The link threshold of 0.60 is fixed while the bank grows, and we have not tested how often it matches the wrong identity beyond the 23 speakers seen here.
 
\textbf{Relation to prior work.} A local diarizer with limited capacity paired with a global linker is the EEND-VC design~\cite{eendvc1}. EEND-GLA~\cite{horiguchi2023gla} trains the linker jointly and has no public checkpoint, and the speaker-tracing buffer~\cite{xue2021stb} carries identity across chunks of a fixed EEND. diart~\cite{diart} is the nearest ancestor, online and overlap-aware, and inspired our overlap-weighted pooling and incremental centroid assignment. We differ in working on a frozen host's persistent slot tracks rather than a re-solved 5\,s window, in deriving cannot-link from concurrent activity rather than from every local speaker, in adding the relink margin and fold-back, and in never altering a frame of activity, which diart's assignment can. The host's arrival-order cache~\cite{streamsortformer} is not an identity store. Its rows separate same- from different-speaker pairs at AUC 0.705 and hold 3.52\,s of audio per slot, so the trunk re-derives identity every chunk. Streaming EEND raises the cap by training~\cite{fseend,lseend}, to 4 and 8 channels in released checkpoints, and Sec.~\ref{ssec:alt} measures that move on this host. Speaker caches serve the same end inside speech LLMs~\cite{gstar,jedisllm}, and practitioners report the problem\footnote{\url{https://github.com/NVIDIA-NeMo/Speech/issues/14546}}.

\section{Experimental setup}
\label{sec:setup}
\textbf{Suite and strata.} One suite of 1418 sessions, 213.5\,h, from eight corpora: Indic-DiarBench~\cite{indicdiarbench} (1164 sessions, 22 Indian languages), Earnings-21~\cite{earnings21}, DISPLACE-2024~\cite{displace2024} and DISPLACE-2026 Hindi and Kannada~\cite{displace2026}, AVA-AVD~\cite{avaavd}, AMI-SDM~\cite{ami2005}, and four manually annotated Supreme Court hearings (Sec.~\ref{ssec:down}). Strata by reference lifetime speaker count: 913 at 4 or fewer, 505 above 4. Of the 505, 406 are Indic-DiarBench. Only 53 exceed eight speakers, most of them Earnings-21 calls (31), AVA-AVD clips (15) plus three of the four hearings, so the many-speaker evidence rests on three corpora. No system receives the count. Every constant was frozen on disjoint train-role corpora (VoxConverse dev, CallHome, DISPLACE-2024 dev, the AMI-SDM train and the ICSI train fold).

\textbf{Contamination.} Sortformer v2.1's card~\cite{sortformerv21card} lists twelve training corpora without giving splits, among them AMI, ICSI, VoxConverse, AISHELL-4, NOTSOFAR, and NIST SRE 2000, which is the CallHome-English diarization set. Seven of our eight corpora are not on it. AMI is, and the card reports on AMI test SDM, so we mark AMI-SDM in-domain. We dropped VoxConverse, AISHELL-4 and ICSI, named in training and absent from the card's results tables, and NOTSOFAR and CallHome-English, which have no above-cap sessions. Restoring all five moves the below-cap delta from $-0.08$ to $+0.07$ and the pooled gain from $-6.1$ to $-5.1$. Nemotron-3~\cite{nemotron3card} trains on DISPLACE-2024 and DISPLACE-2026 Hindi and Kannada, 136 of our sessions; we keep them, and Sec.~\ref{ssec:alt} reports Nemotron-3 with and without them.
 
\textbf{Scoring and reproducibility.} DER is computed with pyannote 4.1 at a 0.25\,s collar, overlap scored, no UEM, one optimal speaker mapping per session, with miss, false alarm and confusion reported separately. Pooling is duration-weighted. Every interval is a paired session bootstrap with 10\,000 resamples. The v2.1 host (117\,M)~\cite{sortformerv21card} runs at its card's 30.4\,s profile (cache 188, FIFO 40, chunk 340, right context 40, update 300 frames of 80\,ms), batch 1, activation threshold 0.5, no post-processing, torch 2.6.0, CUDA 12.4 on A100. The linker runs TitaNet-large on CPU. Rescoring at collar 0 changes no conclusion.
 
\section{Results}
\label{sec:results}
 
\subsection{Lifting the slot cap}
\label{ssec:cap}
\textbf{Headline.} Above the cap the linker takes the frozen host from 36.81 to 22.64 DER (Table~\ref{tab:base}), a 14.17-point cut delivered entirely as confusion, since miss and false alarm are identical by construction. Below the cap it costs $-0.08$, nothing measurable: the paired session bootstrap CI is $[-0.41, +0.24]$.
 
\textbf{The improvement is not uniform.} Earnings-21~\cite{earnings21} (39.6 to 18.1) and the hearings (51.9 to 24.9, Sec. 4.3) carry most of it, while Indic-DiarBench (406 of the 505 above-cap sessions) moves 1.5 over the whole corpus, and within it near-field audio only 0.11 against 8.1 far-field and 9.8 in the wild. Per corpus the linker improves five of eight and costs at most +0.56 on the other three, all below-cap-only sets with nothing to recover.
 
\begin{table}[t]
\centering\small
\setlength{\tabcolsep}{5pt}
\caption{Single-knob ablations, speaker confusion, 1418 sessions. $\Delta$ is variant minus full linker (positive hurts); bold: 95\,\% paired session bootstrap CI excludes zero.
}
\label{tab:abl}
\begin{tabular}{lrrrr}
\toprule
configuration & \lefour & $\Delta$ & \gt{4} & $\Delta$ \\
\midrule
frozen host, no linker & 4.07 & $+$0.08 & 25.29 & \textbf{$+$14.17} \\
\textbf{full linker} & \textbf{3.99} & --- & \textbf{11.12} & --- \\
\midrule
bank $\to$ greedy & 9.24 & \textbf{$+$5.25} & 12.29 & $+$1.17 \\
$-$ fold-back & 5.34 & \textbf{$+$1.35} & 11.39 & \textbf{$+$0.27} \\
fold-back 15\,s $\to$ 7\,s & 4.40 & \textbf{$+$0.41} & 11.05 & $-$0.07 \\
$-$ clean-frame selection & 4.56 & \textbf{$+$0.57} & 12.28 & \textbf{$+$1.16} \\
$-$ cannot-link & 4.24 & \textbf{$+$0.25} & 11.37 & $+$0.25 \\
$-$ relink margin & 3.93 & $-$0.06 & 11.13 & $+$0.01 \\
TitaNet-large $\to$ small & 4.31 & \textbf{$+$0.32} & 16.71 & \textbf{$+$5.59} \\
\bottomrule
\end{tabular}
\end{table}

\textbf{Ablations.} Table~\ref{tab:abl} changes one component at a time. Each variant is replayed from the same host pass over the same regions, so it reports confusion alone. Below the cap the host is already right, so the linker's job is to leave it alone, and the bank and fold-back are what do that: replacing the bank with greedy nearest-centroid assignment adds 5.25 points of confusion there and removing fold-back adds 1.35, while above the cap the same changes cost 1.17, not significant, and 0.27. Above the cap, embeddings matter: the small TitaNet costs 5.59 points and dropping clean-frame selection 1.16. Cannot-link and the relink margin guard rare events, so their pooled effect is small and their above-cap CIs include zero.

\begin{table}[t]
\centering\small
\setlength{\tabcolsep}{3pt}
\caption{Pooled DER / speaker confusion, every system on the same 1418 sessions. Bold: the frozen host plus the linker.}
\label{tab:base}
\begin{tabular}{lcrrr}
\toprule
system & cap & \lefour & \gt{4} & all \\
\midrule
frozen host & 4 & 15.32/4.07 & 36.81/25.29 & 24.50/13.14 \\
\quad + linker & -- & \textbf{15.24/3.99} & \textbf{22.64/11.13} & \textbf{18.41/7.04} \\
\midrule
Ultra (widened) & 8 & 17.23/6.26 & 35.59/20.76 & 25.08/12.46 \\
\quad + linker & -- & 17.01/6.04 & 27.33/12.51 & 21.42/8.80 \\
Nemotron-3$^\ddagger$ & 8 & 12.83/1.59 & 21.86/8.09 & 16.69/4.36 \\
\quad + linker$^\ddagger$ & -- & 13.40/2.15 & 19.74/5.96 & 16.11/3.78 \\
\midrule
diart & 20 & 25.91/13.08 & 41.91/29.03 & 32.75/19.90 \\
LS-EEND & 8 & 35.52/17.87 & 49.42/28.35 & 41.46/22.35 \\
\quad + linker & -- & 31.94/14.30 & 34.36/13.29 & 32.98/13.87 \\
\midrule
pyannote c-1$^\P$ & -- & 16.79/6.57 & 17.72/6.55 & 17.19/6.56 \\
DiariZen$^\P$ & -- & 12.61/2.74 & 16.95/5.63 & 14.47/3.97 \\
\bottomrule
\multicolumn{5}{@{}p{\columnwidth}@{}}{\rule{0pt}{2.4ex}Latency: host and Nemotron-3 30.4\,s, diart 5\,s, LS-EEND 1\,s; the linker adds none. $^\ddagger$\,trains on the three DISPLACE corpora. $^\P$\,offline, trains on AMI.}
\end{tabular}
\end{table}

\textbf{Other streaming hosts.} We also ran the linker, with every constant still frozen on Sortformer train-role data, on two streaming EEND checkpoints from a different model family~\cite{fseend,lseend}: FS-EEND (cap 4) and LS-EEND (cap 8), both conformer encoders with attractor decoders. On the same 1418 sessions, FS-EEND falls from 55.29 to 44.52 and LS-EEND from 41.46 to 32.98 (Table~\ref{tab:base}), again entirely as confusion. Two things limit what this shows. These hosts leave channels empty even below their caps, so the linker's gain there comes from filling empty channels rather than from recycled slots. Their absolute DERs also stay far above this host's, because every released checkpoint of the family targets telephone or close-talk audio.
 
\subsection{Comparison with alternatives}
\label{ssec:alt}

\textbf{Eight slots, two ways.} The alternative to a linker is a host with more slots, and two released checkpoints give us this: Ultra-Sortformer~\cite{ultrasortformer}, its speaker head widened to eight rows and fine-tuned, and NVIDIA's own successor, Nemotron-3~\cite{nemotron3card}, retrained with an 8-slot head. Widening the head did not buy accuracy: Ultra scores 25.08 pooled against the frozen host's 24.50, and the frozen 4-slot host plus the linker (18.41) beats the widened host plus the same linker (21.42), of which 1.25 points are miss and false alarm. Retraining did: with the linker, Nemotron-3 gives the best streaming numbers on the suite, 16.11 pooled and 19.74 above four speakers (Table~\ref{tab:base}). But eight slots is still a concurrency budget: above eight speakers the linker takes Nemotron-3 from 24.69 to 19.31, and below its cap it pays $+$0.57, the tax an always-on linker levies on a host that is already right. Nemotron-3 trains on the three DISPLACE corpora. On the 1282 sessions outside its training list the two hosts are level below the cap (15.32 against 15.48), so its below-cap lead in Table~\ref{tab:base} comes from the 136 sessions it trained on; its lead above the cap holds (21.93 against 36.84, and linked 19.78 against 22.58).

\textbf{Streaming baselines.} Table~\ref{tab:base} compares, on the same sessions, references and scorer, the linked host with diart~\cite{diart} (pyannote segmentation-3.0 plus WeSpeaker embeddings, 5\,s latency, 20-speaker cap) and LS-EEND~\cite{lseend} (streaming EEND, 8-speaker cap, best released checkpoint). A 12-trial search of diart's three thresholds on our train-role split returned its shipped defaults, so that row is both tuned and default; LS-EEND runs as released; our constants were frozen on the same train-role split. The linked 4-slot host leads diart by 10.7 (\lefour) and 19.3 (\gt{4}) points and LS-EEND by 20.3 and 26.8, and the linker lowers LS-EEND itself as well (Sec.~\ref{ssec:cap}).

\textbf{Offline systems.} DiariZen-md-v2~\cite{diarizen} leads the linked 4-slot host on both strata, and pyannote community-1 leads it above the cap and trails it below. Both see the whole file at once, carry no speaker cap, and train on AMI, so this is not like-for-like; it is the bound on what a frozen streaming host plus a causal linker recovers, and our claim is confined to streaming. Linked Nemotron-3 closes most of that gap (16.11 against DiariZen's 14.47).
 
\subsection{Courtroom hearings and downstream transcription}
\label{ssec:down}

\textbf{Reference.} Four Indian Supreme Court hearings (three of about five hours, one of 18 minutes, 5 to 23 speakers, Indian English with romanised Hindi) came with transcripts and speaker identities whose segment timestamps are forced-alignment pseudo-labels. We re-annotated the timing by hand along each recording's own timeline, correcting every boundary against the audio rather than inside the segment's stored window, which comes from the alignment under test and would confirm its own errors; speech the transcript omitted was inserted and spurious segments deleted. One annotator per hearing, two in total; no inter-annotator agreement statistic was computed. To our knowledge this is the first manually time-aligned speaker reference for the domain, and two hearings exceed the cap by a factor of four. The pseudo-reference scores 41.1\,\% DER against the human labels.

\textbf{Diarization.} The host predicts four speakers on every hearing, 51.9 DER pooled; linked, it recovers 20/23, 14/16 and 10/14 identities at 24.9, against 29.8 for linked Nemotron-3, which predicts eight speakers throughout, and 35.7 for offline DiariZen-md-v2~\cite{diarizen}. On the one 5-speaker hearing the linker is the worst system (30.0 against 25.3 for the host), the below-cap cost made visible.

\textbf{Transcription.} Table~\ref{tab:down} asks whether the confusion the linker removes reaches the transcript. Each system's segments are transcribed by one fixed recogniser per corpus with the speaker label passed through, and an oracle row on the reference segmentation is the diarization-free floor. The linker halves the host's concatenated-minimum-permutation word error rate (cpWER) on Earnings-21 and takes it from 82.7 to 24.4 on the hearings, with word diarization error rate (WDER)~\cite{wder} falling in step. Under time-constrained cpWER at a 5\,s collar the hearings move from 114.5 to 62.0 against an oracle of 23.7; the residual above the oracle is segment timing, which the linker inherits from the host and does not alter. DiariZen-md-v2 leads on Earnings-21 and trails on the hearings, the same domain split the DER tables show.

\begin{table}[t]
\centering\small
\setlength{\tabcolsep}{3pt}
\caption{Speaker-attributed transcription, one fixed recogniser per corpus: Parakeet-TDT 0.6B v2 for Earnings-21, an in-house English model for the hearings.}
\label{tab:down}
\begin{tabular}{lrrrrr}
\toprule
 & \multicolumn{2}{c}{Earnings-21} & \multicolumn{3}{c}{courtroom} \\
\cmidrule(lr){2-3}\cmidrule(lr){4-6}
system & cpWER & WDER & cpWER & tcpWER & WDER \\
\midrule
oracle segmentation & 12.3 & 1.3 & 15.9 & 23.7 & 0.9 \\
\midrule
host & 68.2 & 36.1 & 82.7 & 114.5 & 40.1 \\
\textbf{+ linker} & \textbf{34.2} & \textbf{14.0} & \textbf{24.4} & \textbf{62.0} & \textbf{5.1} \\
\midrule
DiariZen-md-v2 & 17.7 & 4.7 & 35.6 & 70.6 & 11.1 \\
\bottomrule
\end{tabular}
\end{table}

\section{Conclusion}
\label{sec:concl}

We showed that the slot count of a streaming diarizer is a limit on concurrency, not on the number of speakers it can follow. When a slot is reused, the host loses the speaker's identity, which can be recovered outside the model. A causal linker that reads only the frozen host's slot posteriors and the audio does this without training and without changing a frame of voice activity, so every change it makes is to speaker confusion. On 1418 sessions from eight corpora it cut DER above the cap from 36.8 to 22.6 and left DER below the cap unchanged, and the same constants transferred to the host's 8-slot successor and to two streaming EEND hosts from another family. Widening this host's head to eight slots gained less than adding the linker to it unchanged; a retrained 8-slot host does better, but still needs the linker above eight speakers.

Two limits bound the
claim. All results are at the host's 30.4\,s profile~\cite{streamsortformer,park2025sortformernovelapproachpermutationresolved}; we make no latency claim. And below the cap the linker can over-count: on Indic it emits more speakers than the reference on 43\,\% of sessions against the host's 22\,\%, at no DER cost.

\section{Compliance with Ethical Standards}

This study used speech corpora (Sec.~\ref{sec:setup}) released for research under their own licences, and publicly available recordings of Supreme Court hearings for which the authors produced the speaker annotations. No new data were collected from human subjects and no ethical approval was required.

\bibliographystyle{IEEEbib}
\bibliography{strings,refs}

\end{document}